# Cellular automata approach to synchronized traffic flow modelling

Junfang Tian[1], Chenqiang Zhu[1], Rui Jiang[2]

[1]Institute of Systems Engineering, College of Management and Economics, Tianjin University, Tianjin 300072, China

[2]MOE Key Laboratory for Urban Transportation Complex Systems Theory and Technology, Beijing Jiaotong University, Beijing 100044, China

Abstract: Cellular automaton (CA) approach is an important theoretical framework for studying complex system behavior and has been widely applied in various research field. CA traffic flow models have the advantage of flexible evolution rules and high computation efficiency. Therefore, CA develops very quickly and has been widely applied in transportation field. In recent two decades, traffic flow study quickly developed, among which "synchronized flow" is perhaps one of the most important concepts and findings. Many new CA models have been proposed in this direction. This paper makes a review of development of CA models, concerning their ability to reproduce synchronized flow as well as traffic breakdown from free flow to synchronized flow. Finally, future directions have been discussed.

## 1. Introduction

Traffic flow studies can be traced back to the 1930's (Greenshields, 1935). For nearly a century, many efforts have been devoted to the studies, which reveals many important and interesting traffic phenomena (Brackstone and McDonald, 1999; Chowdhury et al., 2000; Helbing, 2001; Kerner, 2004, 2009, 2017; Treiber and Kesting, 2013; Saifuzzaman and Zheng, 2014; Zheng, 2014; Jin et al., 2015).

To understand the characteristics and mechanism of traffic flow evolution, many models have been proposed, which can be roughly classified into macroscopic ones and the microscopic ones. Cellular automaton (CA) models are a type of microscopic models. The CA models have the advantage of flexible evolution rules and high computation efficiency. Therefore, ever since the proposal of the classical Nagel-Schreckenberg (NaSch) model (Nagel and Schreckenberg, 1992), CA develops very quickly and has been widely applied.

The previous several papers (e.g., Chowdhury et al., 2000; Maerivoet and Moor, 2005; Knospe, et al. 2004) have made good reviews of the development of CA models. Nevertheless, these papers were published more than ten years ago. In recent two decades, traffic flow study quickly developed, among which "synchronized flow" is perhaps one of the most important concepts and findings. Many new CA models have been proposed in this direction. Therefore, we believe it is time to make another review of development of CA models, concerning the issue of synchronized flow.

The paper is organized as follows. For self-consistency, in following subsections, the CA approach and its application to traffic flow are briefly introduced. Section 2 reviews empirical findings of "synchronized flow" as well as "traffic breakdown from free flow to synchronized". Section 3 discusses the different types of CA models to simulate and explain these phenomena. Conclusion and outlook are given in section 4.

*1.1 The CA approach*

CA is a dynamic system in which spatially-discrete cells have discrete states, and evolve according to spatially-localized discrete-time update rules (Hanson, 2009). In CA, space is divided into many units by regular meshes. Each element in these regular grids is called a cell, which has a finite set of discrete states. All cells follow the same discrete-time local update rules.

The basic idea of CA is to use a large number of simple components, simple links, and simple rules running in parallel in time and space, to simulate complex and rich phenomena (Wolfram, 1982; Jia et al., 2007). CA is an important theoretical framework for studying complex system behavior, and also one prototype of AI (Duan et al., 2012). It has the following advantages:

(i) A large number of cells with local interactions can depict the collective phenomena and evolution dynamics of complex systems.

(ii) Computing is simple and efficient. CA is a highly discrete system, which is manifested in space discrete, time discrete, and state discrete. This main feature makes the computing process significantly simplified. Besides, the computing process is carried out synchronously, which is suitable for parallel computation, and the efficiency is greatly improved.

(iii) Update rules are flexible and intuitive. CA is not determined by strictly defined physical equations or functions. Instead it is defined by a series of evolution rules. The rules can be described in a simple language, which makes it easy to understand.

Up to now, CA has been widely used in social, economic and scientific research fields, including biology (Edelsteinkeshet, 2017), ecology (Hogeweg, 1988), physics (Krug and Spohn, 1988), chemistry (Lodder et al., 1988), transportation science (Chowdhury et al., 2000), computer science (Rosin, 2006), environmental science (Parsons and Fonstad, 2010), society and economy (Goldenberg et al., 2002), financial (Bartolozzi and Thomas, 2004), land use and land cover change (Lauf et al., 2012).

*1.2 Application in traffic flow modelling*

In the CA traffic flow models, the road is classified into cells. Each cell can be empty or occupied by at most one vehicle (except in multi-value CA models, in which each cell can hold more than one vehicle, see e.g., Nishinari and Takahash (2000)). Time is discretized and each time step usually corresponds to 1 s. As a result, velocity and acceleration are also discretized.

This subsection reviews two important CA models. Although the two models cannot reproduce synchronized flow, they are the basics for subsequently developed models. In this review paper, we focus on the car-following behavior on a single circular lane. The lane changing behavior is another important topic and is left for another review paper.

*1.2.1 The NaSch model*

The first CA traffic flow model can be traced back to the Wolfram 184 model (Wolfram, 1982), in which each vehicle moves deterministically. In one time step, vehicles either stay motionless (if cell in front is occupied) or move forward by one cell (if cell in front is empty). Cremer and Ludwig (1986) proposed another CA model. However, these early models did not attract the attention of traffic flow community until the proposal of NaSch model.

The parallel update rules in the NaSch model are as follows.

Step 1. Acceleration:
$$\tilde{v} = \min(v + a, v_{max}).$$

Step 2. Deceleration:
$$\tilde{v} = \min(\tilde{v}, d).$$

Step 3. Randomization:
$$v' = \begin{cases} \min(\tilde{v} - b, 0), & \text{if } r < p, \\ \tilde{v}, & \text{otherwise.} \end{cases}$$

Step 4. Vehicle movement:
$$x' = x + v'.$$

Here $x$ ($x'$) and $v$ ($v'$) denote speeds and positions at the current and next time step. $x_l$ is the position of the leading vehicle. $d = x_l - x - L_{veh}$ is space gap between the two vehicles and $L_{veh}$ is vehicle length. $v_{max}$ is maximum speed of vehicles. $a$ and $b$ are acceleration and randomization deceleration, respectively, which are usually set to be 1. $p$ is randomization probability.

Fig. 1 shows the fundamental diagram of the NaSch model on a circular road with length 1000 $L_{cell}$. The parameters are shown in Table 1. Two different traffic states are classified. When the density is smaller than the critical density $\rho_c$, the traffic is in free flow, see Fig. 2(a). When $\rho > \rho_c$, the jams appear spontaneously, see Fig. 2(b).

Table 1. Parameter values of the NaSch model. Taken from Nagel and Schreckenberg (1992).

| Parameters | $L_{cell}$ | $L_{veh}$ | $v_{max}$ | $a$ | $b$ | $p$ |
|---|---|---|---|---|---|---|
| Units | m | $L_{cell}$ | $L_{cell}$/s | $L_{cell}$/s$^2$ | $L_{cell}$/s$^2$ | - |
| Value | 7.5 | 1 | 5 | 1 | 1 | 0.3 |

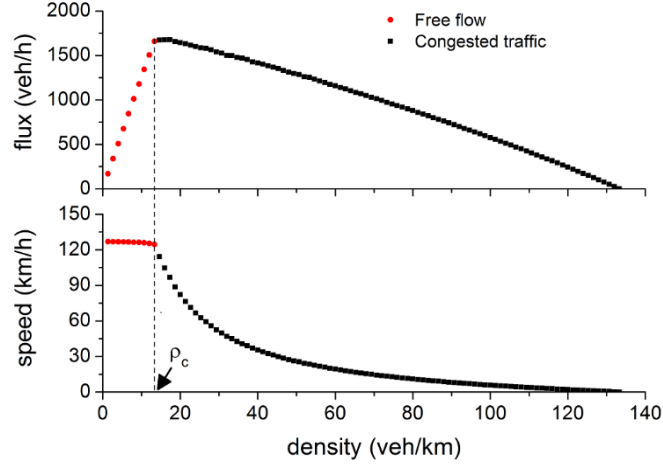

Fig. 1. The fundamental diagram of the NaSch model on a circular road. This figure describes the averaged flow as a function of the global density (number of vehicles divided by the circumference). The averaged flow is product of the global density and the average speed of all vehicles on the road.

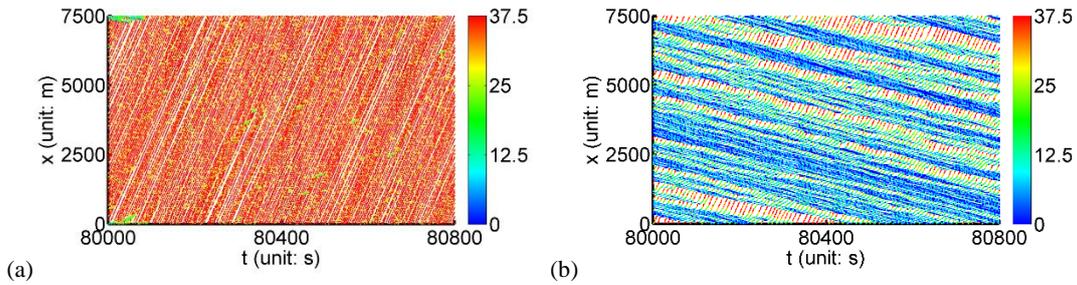

Fig. 2. The spatiotemporal diagrams of NaSch model on the circular road. (a) $\rho$ = 13veh/km; (b) $\rho$ = 53veh/km. The color bar indicates speed (m/s).

As recently revealed, stochastic factors play a nontrivial role in traffic flow dynamics. Traffic flow instability might be due to cumulative effect of stochastic factors (e.g., Jiang et al., 2014, 2015, 2018; Laval et al., 2014; Treiber and Kesting, 2017). Randomization in NaSch model characterizes the stochastic factors, which enables the model to depict the spontaneous formation of jams in congested flow. This feature makes NaSch model a milestone in traffic flow studies, which has been cited 2401 times according to Web of Science (accessed on 23/09/2018). Moreover, the NaSch model has been widely used in transportation application. For example, it has been used as microscopic simulator in the TRANSIMS developed at Los Alams National Laboratory (Nagel and Rickert, 2001) and in urban traffic simulation in Dallas/Fort-Worth area, (Rickert and Nagel, 1997).

### 1.2.2 The Slow-to-start models

In the NaSch model, the transition from free flow to congested flow is not of first-order. To depict the first order transition in traffic flow, the slow-to-start rules are introduced (Takayasu and Takayasu, 1993; Benjamin et al., 1996; Fukui et al., 1997; Schadschneider et al., 1997; Barlovic et al., 1998). Here we review the velocity-dependent-randomization (VDR) rule (Barlovic et al., 1998), which is the simplest one. In the VDR model, the only difference from the NaSch model is that the randomization is not a constant, it depends on the velocity as follows,

$$p(v) = \begin{cases} p_0, & \text{if } v=0, \\ p, & \text{otherwise.} \end{cases}$$

Fig. 3 shows the fundamental diagram of the VDR model, in which there are two critical densities. Below $\rho_{c1}$, the traffic is in free flow; above $\rho_{c2}$, the traffic is in jam. When the density is in the range $\rho_{c1} < \rho < \rho_{c2}$, the traffic flow is metastable. When starting from homogeneous traffic, free flow is maintained, see Fig. 4(a). When starting from megajam, the traffic is in jam. Moreover, the first-order transition from free flow to jam could appear spontaneously, see Fig. 4(b).

Table 2. Parameter values of the VDR model taken from Barlovic et al. (1998).

| Parameters | $L_{cell}$ | $L_{veh}$ | $v_{max}$ | $a$ | $b$ | $p_0$ | $p$ |
|---|---|---|---|---|---|---|---|
| Units | m | $L_{cell}$ | $L_{cell}/s$ | $L_{cell}/s^2$ | $L_{cell}/s^2$ | - | - |
| Value | 7.5 | 1 | 5 | 1 | 1 | 0.75 | 1/64 |

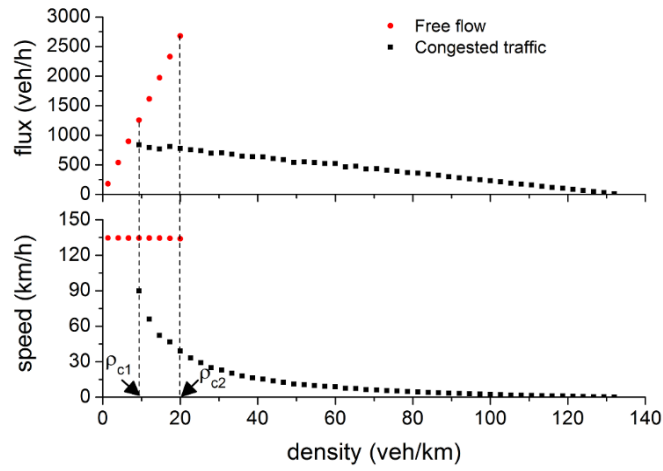

Fig. 3. The fundamental diagram of the VDR model on a circular road. The parameters are shown in Table **2**.

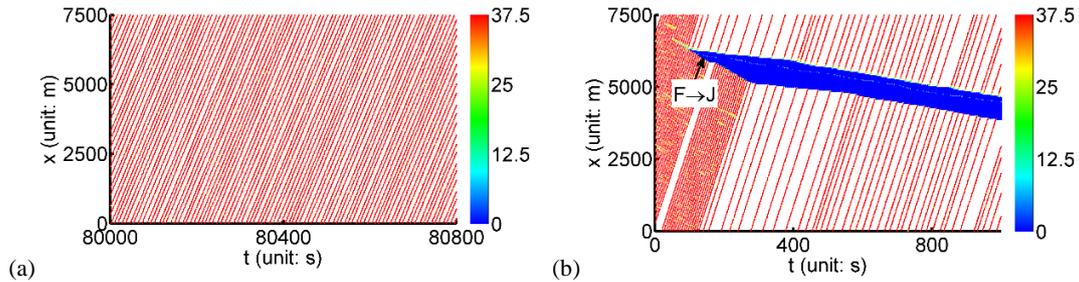

Fig. 4. The spatiotemporal diagrams of VDR model on the circular road. (a) $\rho$ = 15veh/km, start from homogeneous initial condition; (b) $\rho$ = 20veh/km.

## 2. Empirical findings of synchronized flow

*2.1 Synchronized traffic flow*

Based on a long-term empirical data analysis, Kerner introduced the three-phase traffic theory

(Kerner, 1998, 2004, 2009)[1], which classifies the congested traffic into the synchronized flow (S) and the wide moving jams (J). In congested traffic flow, when speed is not small (or the density is not large), jam will not appear spontaneously. This "synchronized flow" is stable (see Fig. 5). With decrease of speed (or the increase of density), synchronized flow becomes unstable and jam will appear spontaneously (see Fig. 6).

Apart from USA, China, synchronized traffic flow has been reported among different countries, such as the UK (see Rehborn, et al. 2011), and Germany (see Kerner 2004); Iran (see Kouhi et al., 2018); Singapore (see Yang et al., 2016).

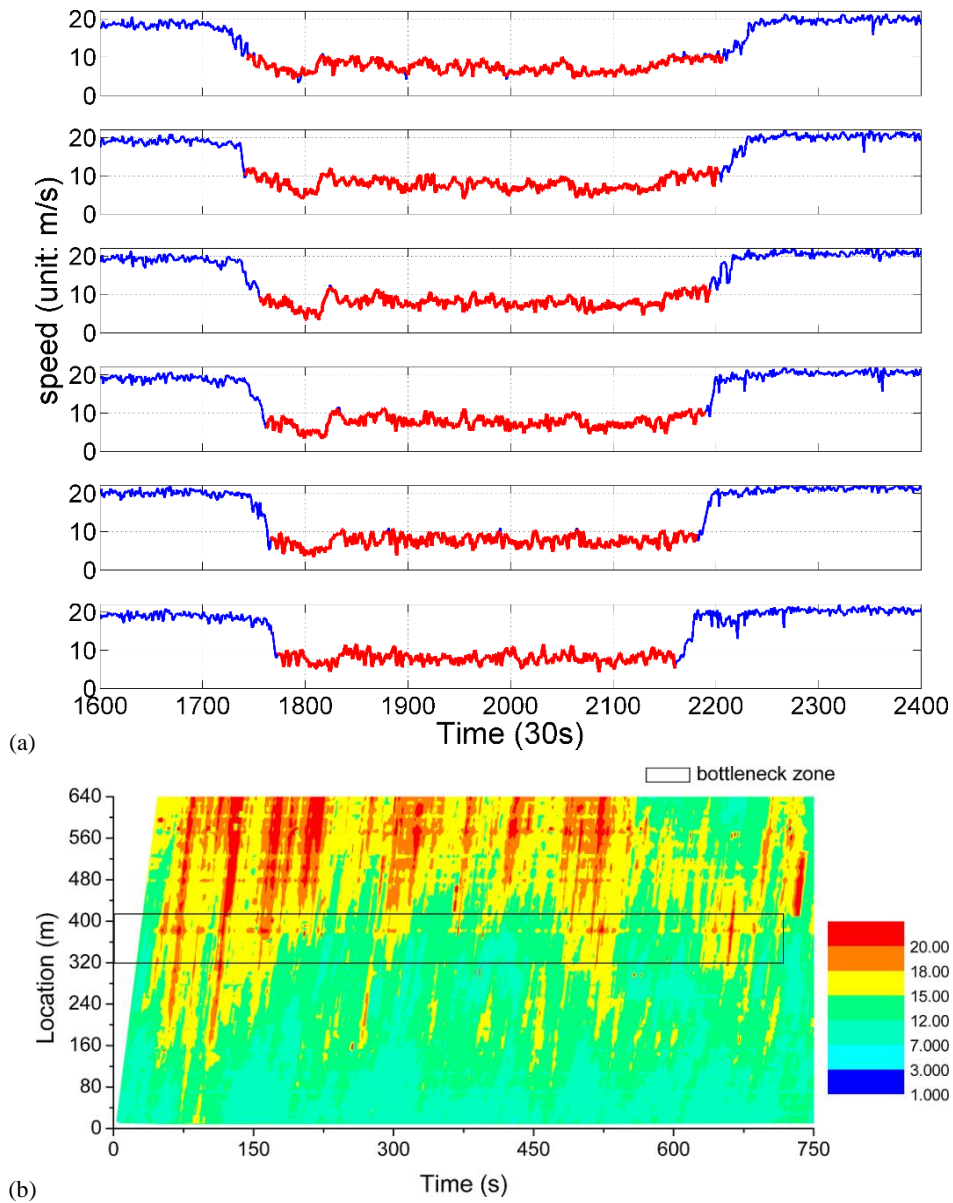

(a)

(b)

---

[1] We would like to mention that there is controversy about "three-phase traffic theory". However, the existence of synchronized flow is not controversial. For example, Helbing et al. published a paper entitled "criticism of three-phase traffic theory" (Helbing and Schoenhof, 2009), but they also showed the empirical data of synchronized flow (Helbing et al., 2009).

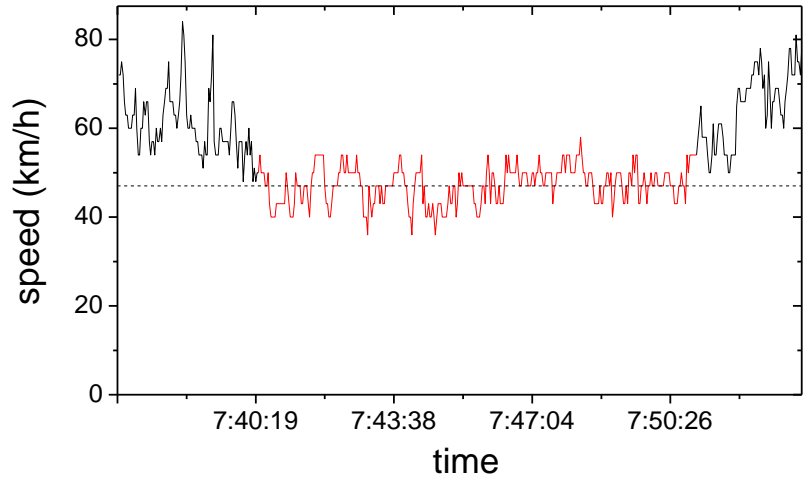

(c)

Fig. 5. (a) The speed time series at different detector stations (unit: m/s) in I80. Panels from up to bottom correspond to the stations 7, 6,5,4,3 and 1 respectively. The direction of traffic flow is from station 7 to 1. The data of station 2 is missing. The red denotes synchronized flow. (b) US 101; upstream of the bottleneck the traffic flow is synchronized flow (sometimes alternating with free flow). (c) Nanjing airport. The red denotes synchronized flow.

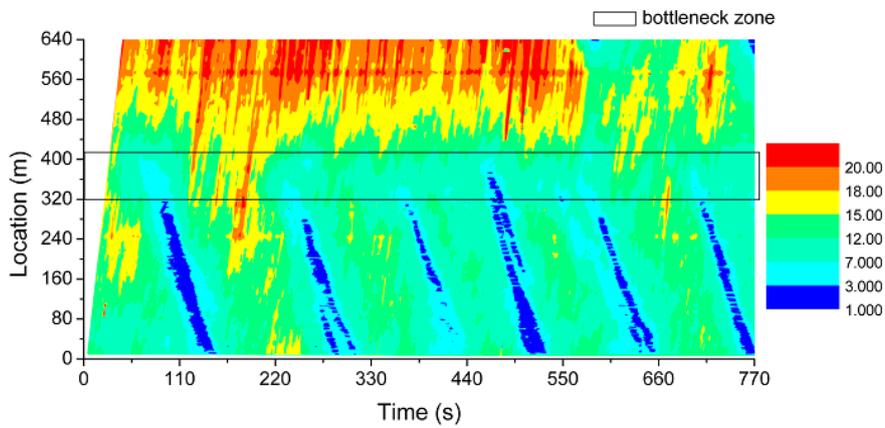

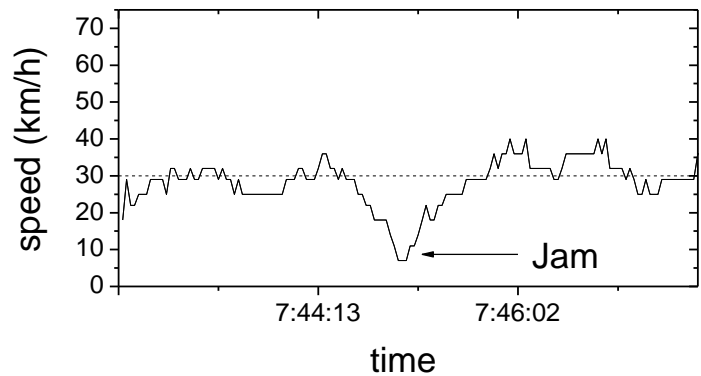

Fig. 6. (a) US101 leftmost lane. The blue stripes denote jams. (b) Nanjing airport.

*2.2 Traffic breakdown from free flow to synchronized flow*

Traffic breakdown from free flow to synchronized flow is characterized by a sharp decrease in speed, an abrupt increase in density, and in particular a plummeting drop in capacity (Banks 1991, Agyemang-Duah and Hall, 1991; Cassidy and Bertini, 1999). Traffic breakdown often happens at flow the non-conservation bottlenecks, such as ramps and lane drops. Athol and Bullen (1973), Elefteriadou et al. (1995), Lorenz and Elefteriadou (2001) and Zhang and Levinson (2004) reported that breakdown has the probabilistic nature, i.e. its occurrence probability increases with increasing flow rate.

At the flow conservation bottlenecks, traffic breakdown also has been observed, such as the tunnels, vertical alignment sags, S curves (Koshi, 1986; Koshi et al. 1992). Related observations were discovered in many places, see e.g., Edie and Foote (1958), Newell (1959), Iwasaki (1991), Tadaki, et al. (2002), Brilon and Bressler (2004), Chung et al. (2007).

Furthermore, traffic breakdown was also observed on single lane roads. For example, Shiomi et al. (2011) reported the breakdown on a single lane road where the tunnel acts as the bottleneck. Jin et al. (2013) investigated the empirical data on a single-lane highway section and found that spontaneous sudden drops of speed can happen in the platoon when the speed of the leading vehicle is quite high (~70 km/h), which shows that traffic breakdown may be the phase transition from free flow to synchronized flow on a single-lane road.

## 3. Cellular automaton models producing synchronized traffic flow

To reproduce the synchronized flow, many CA models have been proposed. Instead of reviewing these models one by one, we classify the models into different types, concerning their main characteristics. We review one or several typical models for each type.

*3.1 Models considering speed adaptation effect explicitly*

The Kerner-Klenov-Wolf (KKW, Kerner et al. 2002) model is one of the first CA models (the other is the comfortable driving model in section 4.3) within the framework of three-phase traffic theory, in which the 2D region of steady state and the speed adaptation effect are explicitly considered. The update rules consist of the deterministic and stochastic rules.

Step 1. Dynamical part of all KKW models:

$$\tilde{v} = \max\left(0, \min\left(v_{\max}, v_s, v_c\right)\right).$$

where,

$$v_s = d.$$

$$v_c = \begin{cases} v + a & \text{for } x_l - x > G(v), \\ v + a\,\text{sign}(\Delta v) & \text{for } x_l - x \leq G(v). \end{cases}$$

Step 2. Stochastic part of all KKW models:

$$v' = \max\left(0, \min\left(\tilde{v} + a\eta, v + a, v_{max}, v_s\right)\right).$$

where,

$$\eta = \begin{cases} -1, & \text{if } r < p_b, \\ 1, & \text{if } p_b \leq r < p_b + p_a, \\ 0, & \text{otherwise.} \end{cases}$$

$$p_b = \begin{cases} p_0, & \text{if } v = 0, \\ p, & \text{if } v > 0. \end{cases}$$

$$p_a = \begin{cases} p_{a1}, & \text{if } v < v_p, \\ p_{a2}, & \text{if } v \geq v_p. \end{cases}$$

Step 3. Vehicle movement:

$$x' = x + v'$$

Here $\Delta v = v_l - v$ is the speed difference and $v_l$ is the speed of the leading vehicle. $v_s$ is the safe speed that must not be exceeded to avoid collisions. $v_c$ describes the rule of "speed change", which means the acceleration behavior depends on whether the leading vehicle is within a synchronized space gap. The synchronized space gap can be defined as linear form: $G(v) = L_{veh} + kv$ or nonlinear form: $G(v) = L_{veh} + v + \beta v/(2a)$, in which $k$ and $\beta$ are the positive parameters. The sign function is defined as

$$\text{sign}(\Delta v) = \begin{cases} -1 & \text{if } \Delta v < 0, \\ 0 & \text{if } \Delta v = 0, \\ 1 & \text{if } \Delta v > 0. \end{cases}$$

The steady state corresponds to a 2D region bounded by $q = \rho v_s = (1-\rho L_{veh})$ (U: related to safe speed), $q = \rho v_{max}$ (F: related to maximum speed) and $q = (1-\rho L_{veh})/k$ (L: related to the linear form of synchronized space gap), see Fig. 7. The rule $v_c = v + a \, \text{sign}(\Delta v)$ corresponds to the speed adaptation effect, which occurs when the traffic state is in the 2D region.

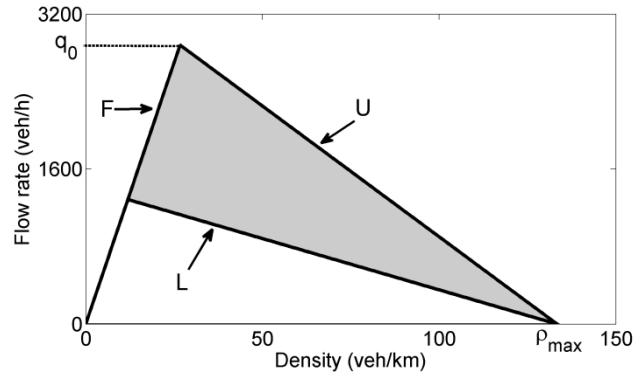

Fig. 7. The 2D region of KKW model.

Fig. 8 shows simulation results of the fundamental diagram of the KKW model on a circular road with length $L=7500$m. Each cell corresponds to $L_{cell} = 0.5$ m. The parameter values are shown in Table 3. One can see that three density ranges are classified by two critical densities. When the density $\rho < \rho_{c1}$, the traffic flow is in free flow state, see Fig. 9(a). When the density is larger than $\rho_{c1}$, traffic flow is a coexistence of free flow and jams if initially starting from a megajam, see Fig. 9(b). When the density $\rho_{c1} < \rho < \rho_{c2}$, the traffic flow will be in synchronized flow if starting from a homogeneous configuration, see Fig. 9(c). When $\rho > \rho_{c2}$, synchronized flow cannot be maintained even if starting from homogeneous configuration. Jams will emerge spontaneously, see Fig. 9(d). We also notice that in the simulation results, the flow rate in synchronized flow changes non-monotonically, which might be related to the choice of parameters.

Table 3. Parameter values of the KKW model taken from Kerner et al. (2002).

| Parameters | $L_{cell}$ | $L_{veh}$ | $v_{max}$ | $a$ | $k$ | $p$ | $p_0$ | $p_{a1}$ | $p_{a2}$ | $v_p$ |
|---|---|---|---|---|---|---|---|---|---|---|
| Units | m | $L_{cell}$ | $L_{cell}/s$ | $L_{cell}/s^2$ | - | - | - | - | - | $L_{cell}/s$ |
| Value | 0.5 | 15 | 60 | 1 | 2.55 | 0.04 | 0.425 | 0.2 | 0.052 | 28 |

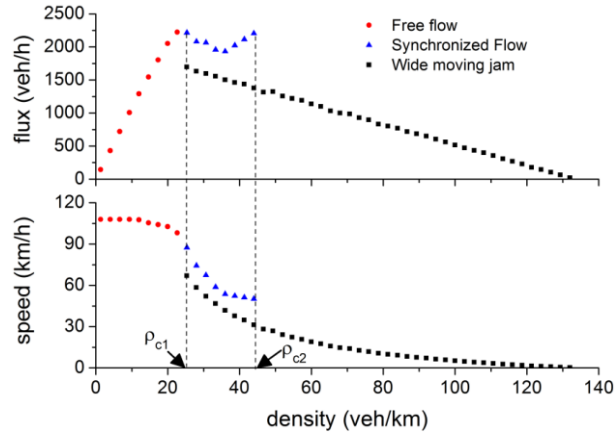

Fig. 8. The fundamental diagram of the KKW model on a circular road. The parameters are shown in Table 3.

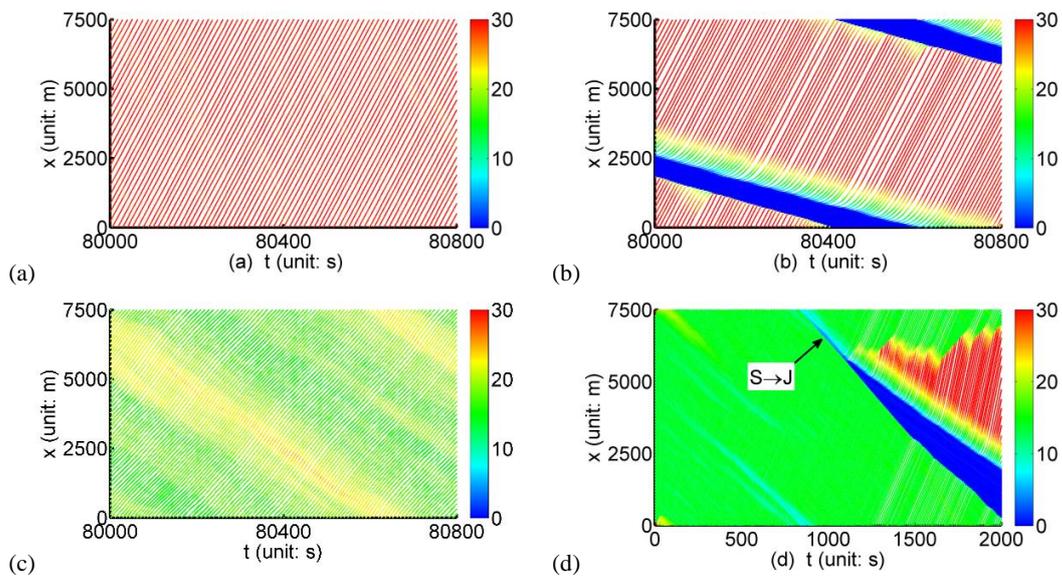

Fig. 9. The spatiotemporal diagrams of KKW model on the circular road. (a) $\rho = 15\,\text{veh/km}$; (b) $\rho = 31\,\text{veh/km}$; (c) $\rho = 31\,\text{veh/km}$; (d) $\rho = 47\,\text{veh/km}$.

## 3.2 Generalized NaSch models

Models of this class (such as Gao et al. 2007, 2009; Zhao et al. 2009; Jia et al. 2011; Tian et al. 2012a, 2012b, 2015, 2016) are all based on the following generalized NaSch rules.

Step 1. Acceleration/deceleration:
$$\tilde{v} = \min\left(v + \hat{a}, \hat{d}, \hat{v}\right).$$

Step 2. Randomization with probability $\hat{p}$:
$$v' = \begin{cases} \min\left(\tilde{v} - \hat{v}_r, 0\right), & \text{if } r < \hat{p}, \\ \tilde{v}, & \text{otherwise.} \end{cases}$$

Step 3. Vehicle movement:
$$x' = x + v'.$$

In different models, the formulations of $\hat{a}$, $\hat{d}$, $\hat{v}$, $\hat{v}_r$, and $\hat{p}$ are different.

### 3.2.1 The model of Gao et al.

In the model of Gao et al. (2007), the formulations of $\hat{a}$, $\hat{d}$, $\hat{v}$, $\hat{v}_r$, and $\hat{p}$ are as follows.

(i) $\hat{a} = a$

(ii) $\hat{d} = d$

(iii) $\hat{v} = v_{\max}$

(iv) $\hat{v}_r = \begin{cases} a & \text{when } t_{st} \geq t_c, \\ \begin{cases} b_- & \text{if } v < v_l \\ b_0 & \text{if } v = v_l \\ b_+ & \text{if } v > v_l \end{cases} & \text{otherwise.} \end{cases}$

(v) $\hat{p} = \begin{cases} p_0 & \text{when } t_{st} \geq t_c, \\ p_d & \text{when } t_{st} < t_c. \end{cases}$

Here $t_{st}$ denotes the time that a vehicle stops,
$$t'_{st} = \begin{cases} t_{st} + 1 & \text{if } v = 0, \\ 0 & \text{if } v > 0. \end{cases}$$

where, $t_c$ is a slow-to-start parameter, $b_-$, $b_0$ and $b_+$ are the randomization decelerations corresponding to different velocity difference conditions. Note that the speed adaptation effect is implicitly considered in the formulation of $\hat{v}_r$.

Fig. 10 shows simulation results of the fundamental diagram of the Gao model, which is similar to that of KKW model, except that the synchronized flow branch monotonically decreases with the density. This model could reproduce the synchronized flow, the jam and the S→J transition, see Fig. 11. However, it fails to reproduce the F→S transition.

Table 4. Parameter values of Gao model, taken from Gao et al. (2007).

| Parameters | $L_{cell}$ | $L_{veh}$ | $v_{max}$ | $t_c$ | $p_d$ | $p_0$ | $a$ | $b_-$ | $b_0$ | $b_+$ |
|---|---|---|---|---|---|---|---|---|---|---|
| Units | m | $L_{cell}$ | $L_{cell}/s$ | s | - | - | $L_{cell}/s^2$ | $L_{cell}/s^2$ | $L_{cell}/s^2$ | $L_{cell}/s^2$ |
| Value | 1.5 | 5 | 25 | 7 | 0.3 | 0.6 | 2 | 1 | 2 | 5 |

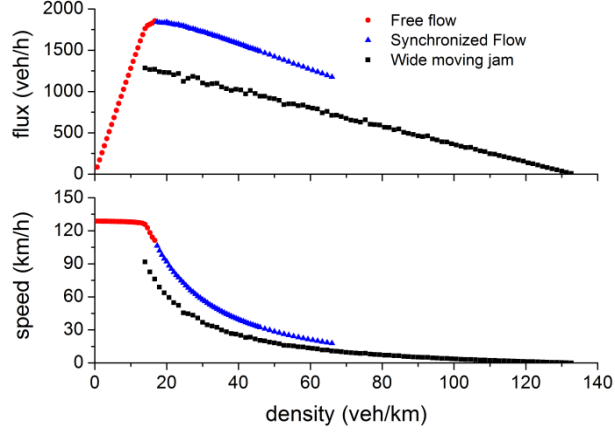

Fig. 10. The fundamental diagram of Gao model on the circular road with length L=7500m. The parameters are shown in Table **4**.

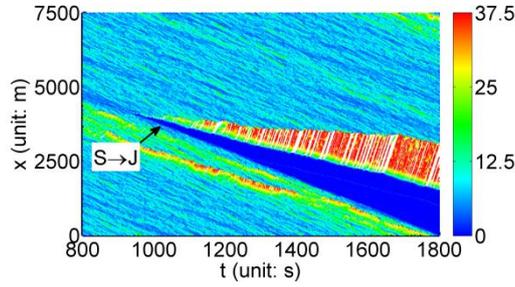

Fig. 11. S→J transition in the Gao model, $\rho$ = 69veh/km.

To overcome this deficiency, Gao et al. (2009) later proposed an improved model, where $\hat{v}_r$, and $\hat{p}$ are changed into.

(iv) $\hat{v}_r = \begin{cases} a & \text{when } t_{st} \geq t_c, \\ \begin{cases} b_- & \text{if } v < v_l \\ b_0 & \text{if } v = v_l \\ b_+ & \text{if } v > v_l \end{cases} & \text{when } t_{st} < t_c \text{ and } d < D \\ b_s & \text{otherwise.} \end{cases}$

(v) $\hat{p} = \begin{cases} p_0 & \text{when } t_{st} \geq t_c, \\ p_d & \text{when } t_{st} < t_c \text{ and } d \leq D, \\ p_s & \text{when } t_{st} < t_c \text{ and } d > D. \end{cases}$

Here $D$ is an effective operating distance in velocity adaptation mechanism. Fig. 12 shows simulation results of the fundamental diagram of the improved Gao model. This model could reproduce the synchronized flow, the F→S and the S→J transition, see Fig. 13.

We would like to mention that in the synchronized flow branch, the traffic flow is a coexistence of free flow and synchronized flow, see Fig. 13(a). With the increase of density, the free flow region shrinks and finally vanishes, see Fig. 14(a). We also point out that in the synchronized flow in the improved Gao model, the speed fluctuation of vehicles is too strong, which seems unrealistic, see Fig. 14(b).

Table 5. Parameter values of improved Gao model, taken from Gao et al. (2009).

| Parameters | $L_{cell}$ | $L_{veh}$ | $v_{max}$ | $t_c$ | $p_d$ | $p_0$ | $p_s$ |
|---|---|---|---|---|---|---|---|
| Units | m | $L_{cell}$ | $L_{cell}/s$ | s | - | - | - |
| Value | 1.5 | 5 | 25 | 6 | 0.18 | 0.5 | 0.08 |
| Parameters | $a$ | $b_-$ | $b_s$ | $b_0$ | $b_+$ | $D$ | |
| Units | $L_{cell}/s^2$ | $L_{cell}/s^2$ | $L_{cell}/s^2$ | $L_{cell}/s^2$ | $L_{cell}/s^2$ | $L_{cell}$ | |
| Value | 2 | 1 | 1 | 2 | 5 | 23 | |

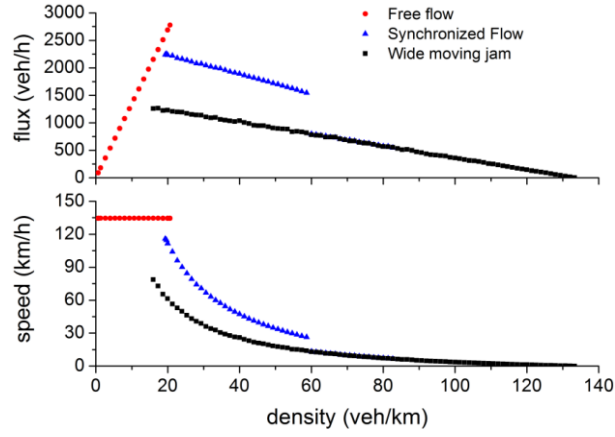

Fig. 12. The fundamental diagram of improved Gao model on the circular road with length L=7500m. The parameters are shown in Table **5**.

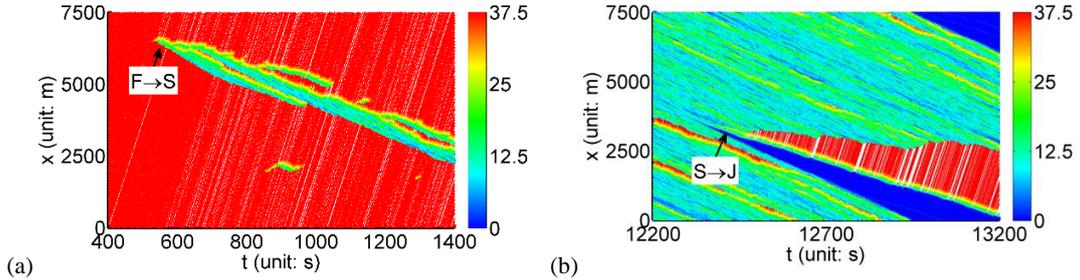

Fig. 13. The simulation results of the spatiotemporal diagram. (a) $\rho$=20veh/km; (b) $\rho$=59veh/km.

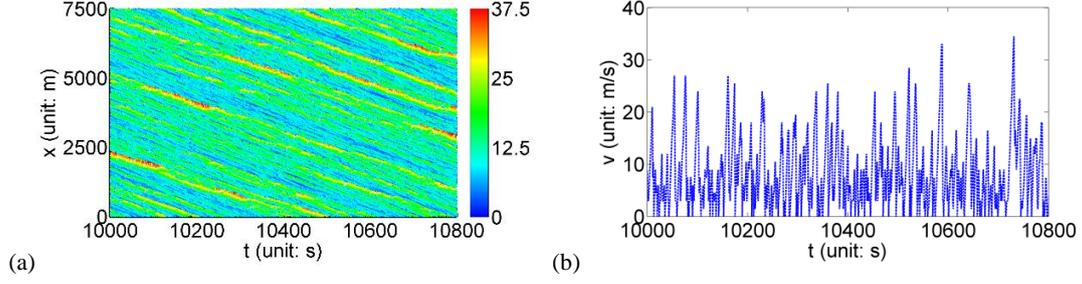

Fig. 14. The simulation results of spatiotemporal diagram. (a) $\rho$=55veh/km; (b) The time series of velocity of a single vehicle.

*3.2.2 Two-state model and its improved version*

In the two-state model (TS model, Tian et al., 2015), the formulations of $\hat{a}$, $\hat{d}$, $\hat{v}$, $\hat{v}_r$, and $\hat{p}$ are as follows.

(i) $\hat{a} = a$

(i) $\hat{d} = d + \max(v_{anti} - g_{safety}, 0)$

(ii) $\hat{v} = v_{max}$

(iii) $\hat{v}_r = \begin{cases} a & \text{if } v \leq d_{anti}/T, \\ b_{defense} & \text{otherwise.} \end{cases}$

(iv) $\hat{p} = \begin{cases} p_b & \text{if } v = 0 \text{ and } t_{st} > t_c, \\ p_c & \text{else if } 0 \leq v \leq d_{anti}/T, \\ p_a & \text{otherwise.} \end{cases}$

Here $v_{anti} = \min(d_l, v_l + a, v_{max})$ is the anticipated speed of the leading vehicle, in which $d_l$ is the space gap of the leading vehicle. The effectiveness of the anticipation is controlled by $g_{safety}$. Accidents are avoided only if $\hat{v}_r \geq g_{safety}$.

The TS model has considered two driver states: a defensive one, and a normal one. The defensive state is activated if $v > d_{anti}/T$. Otherwise the drivers are assumed to be in the normal driving state.

Fig. 15 shows simulation results of the fundamental diagram of the two-state model. One can see that the model can reproduce the free flow, the synchronized flow, the jam, and the F→S (Fig. 16(a)) and S→J transitions (Fig. 16(b)).

Table 6. Parameter values of TS model, taken from Tian et al. (2015).

| Parameters | $L_{cell}$ | $L_{veh}$ | $v_{max}$ | $T$ | $p_a$ | $p_b$ | $p_c$ | $a$ | $b_{defense}$ | $g_{safety}$ | $t_c$ |
|---|---|---|---|---|---|---|---|---|---|---|---|
| Units | m | $L_{cell}$ | $L_{cell}$/s | s | - | - | - | $L_{cell}$/s² | $L_{cell}$/s² | $L_{cell}$ | s |
| Value | 7.5 | 1 | 5 | 1.8 | 0.95 | 0.55 | 0.1 | 1 | 1 | 2 | 8 |

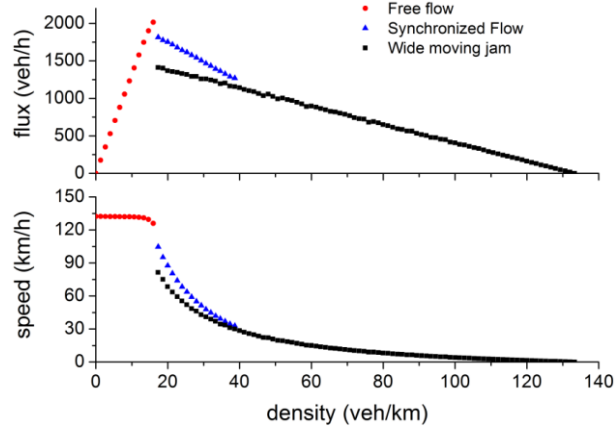

Fig. 15.The fundamental diagram of the TS model on a circular road with length L=3000m. The parameters are shown in Table 6.

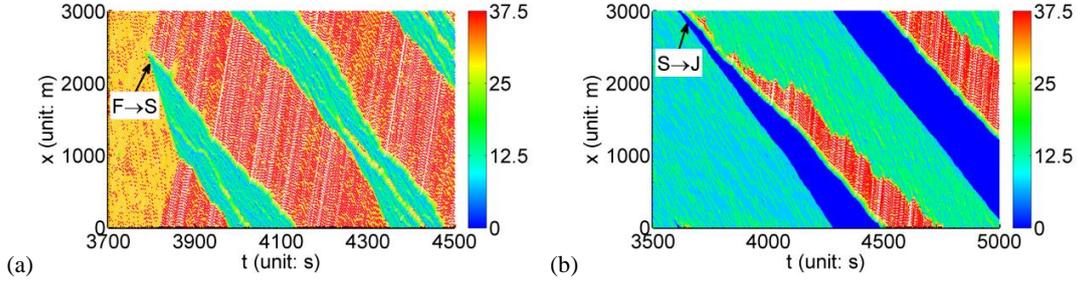

Fig. 16. The spatiotemporal diagrams of TS model with the density $\rho$=25veh/km (a) and (b) $\rho$=67veh/km.

However, in the TS model, the growth of oscillate magnitudes of vehicles in a car platoon are much faster than that reported in the experiment. For more details, one may refer to Tian et al. (2016). To make the two-state model more realistic, the safe speed $v_{safe}$, and the logistic function for the randomization probability $p_{defense}$ are introduced into the Two-state model. The two new variables are defined as follows:

$$v_{safe} = \left[ -b_{max} + \sqrt{b_{max}^2 + v_l^2 + 2b_{max}d_n} \right].$$

$$p_{defense} = p_c + \frac{p_a}{1 + e^{\alpha(v_c - v_n)}}.$$

where $\alpha$ and $v_c$ are the steepness and midpoint of the logistic function, respectively. $v_l$ is the speed of the preceding car, $b_{max}$ is the maximum deceleration of the car. The round function $[x]$ returns the integer nearest to $x$. Since safe speed is introduced, the model is named as the two-state model with safe speed (TSS model, Tian et al., 2016). In the TSS model, the formulations of $\hat{v}$, $\hat{v}_r$, and $\hat{p}$ change into:

$$\hat{v} = \min(v_{max}, v_{safe}).$$

$$\hat{v}_r = \begin{cases} a & \text{if } v < b_{defense} + \lfloor d_{anti}/T \rfloor, \\ b_{defense} & \text{otherwise.} \end{cases}$$

$$\hat{p} = \begin{cases} p_b & \text{if } v = 0, \\ p_c & \text{else if } 0 < v \leq d_{anti}/T, \\ p_{defense} & \text{otherwise.} \end{cases}$$

Here $\lfloor x \rfloor$ returns the maximum integer no greater than $x$. Fig. 17 shows the simulation results of the fundamental diagram of the improved two-state model. One can see that the model can well reproduce the three phases of traffic flow. Fig. 18 shows that the both the F→S and the S→J transition can be well depicted. Moreover, the growth of oscillate magnitudes of vehicles in a car platoon are consistent with that reported in the experiment (not shown here, see Tian et al. (2016)). Note that in the improved two-state model, the synchronized flow could coexist with free flow (see Fig. 18(a)), and traffic oscillations exist in the synchronized flow (see Fig. 19).

Table 7. Parameter values of TSS model taken from Tian et al. (2016).

| Parameters | $L_{cell}$ | $L_{veh}$ | $v_{max}$ | $T$ | $p_a$ | $p_b$ | $p_c$ |
|---|---|---|---|---|---|---|---|
| Units | $m$ | $L_{cell}$ | $L_{cell}/s$ | $s$ | - | - | - |
| Value | 0.5 | 15 | 60 | 1.8 | 0.85 | 0.52 | 0.1 |
| Parameters | $a$ | $b_{max}$ | $b_{defense}$ | $g_{safety}$ | $v_c$ | $\alpha$ | |
| Units | $L_{cell}/s^2$ | $L_{cell}/s^2$ | $L_{cell}/s^2$ | $L_{cell}$ | $L_{cell}/s$ | $s/L_{cell}$ | |
| Value | 1 | 7 | 2 | 20 | 30 | 10 | |

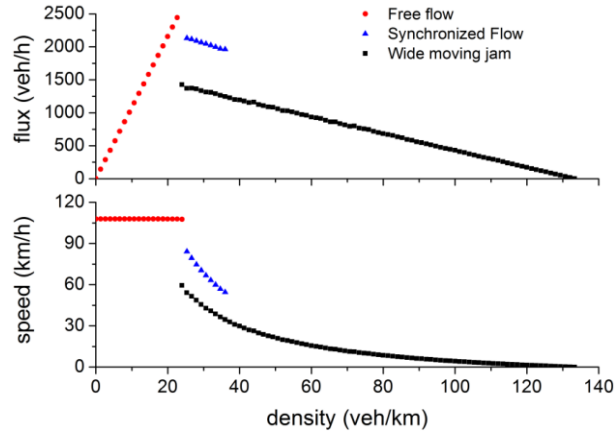

Fig. 17. The fundamental diagram of the improved two-state model on a circular road with length 3000m. The parameters are shown in Table 7.

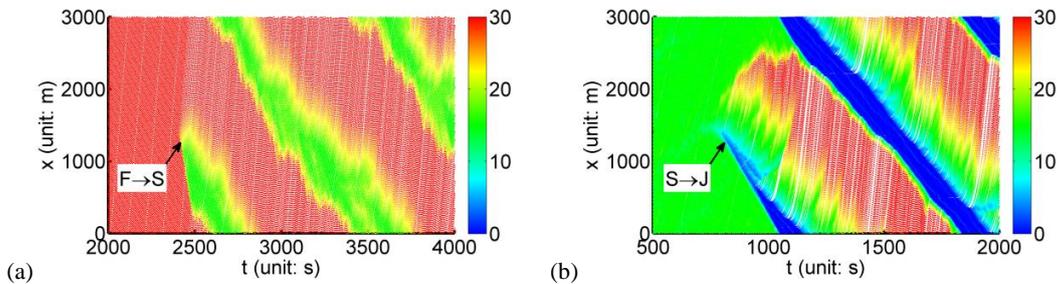

Fig. 18. The spatiotemporal diagrams of improved two-state model on the circular road. (a) $\rho=24$veh/km; (b) $\rho=38$veh/km.

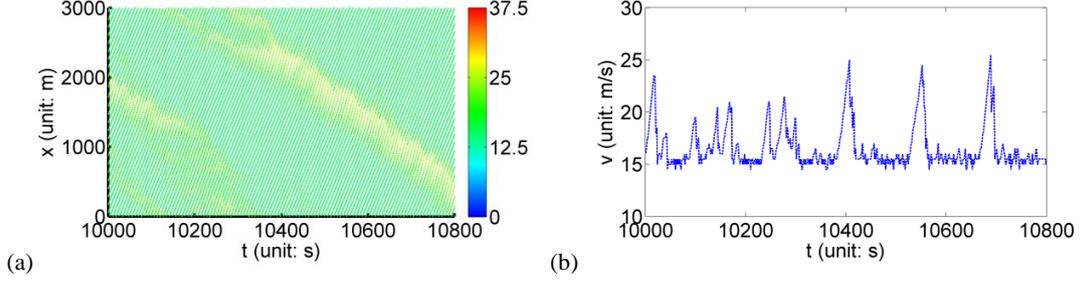

Fig. 19. The simulation results of spatiotemporal diagram. (a) $\rho$=33veh/km; (b) The time series of velocity of a single vehicle.

### 3.3. Brake light models

The class of brake light models also belongs to the generalized NaSch models. Since the status of brake light[2] is considered, we review these models separately in this subsection.

### 3.3.1 The comfortable driving model

Knospe et al. (2000) proposed the comfortable driving (CD) model that considers the desire of the drivers for smooth and comfortable driving. In this model, the formulations of $\hat{a}$, $\hat{d}$, $\hat{v}$, $\hat{v}_r$, and $\hat{p}$ are as follows.

(i) $\hat{a} = \begin{cases} a, & \text{if } (s_l = 0 \text{ and } s = 0) \text{ or } t_h \geq t_s, \\ 0, & \text{otherwise.} \end{cases}$

(ii) $\hat{d} = d + \max(v_{\text{anti}} - g_{\text{safety}}, 0)$.

(iii) $\hat{v} = v_{\max}$

(iv) $\hat{v}_r = b$

(v) $\hat{p} = \begin{cases} p_b, & \text{if } s_l = 1 \text{ and } t_h < t_s, \\ p_0, & \text{if } v_n = 0, \\ p_d, & \text{otherwise.} \end{cases}$

Here $v_{\text{anti}} = \min(d_l, v_l)$. $s$ and $s_l$ are the status of the brake light of the vehicle and its leader ($s = 1(0)$ means the brake lights is on (off)). $t_h = d/v$ is the time gap. $t_s$ is the safe time headway, which is defined as $t_s = \min(v, h)$. The slow-to-start rule is applied to simulate the wide moving jams in which standing vehicles in the previous time step have a higher randomization probability $p_0$ than the moving vehicles with $p_d$. Moreover, $b \geq a$ should be satisfied to simulate the wide moving jams. The brake light state $s'$ in the next time step is determined as follows.

$s' = \begin{cases} 1, & \text{if } \tilde{v} < v \text{ or } (p = p_b \text{ and } v' < v), \\ 0, & \text{otherwise.} \end{cases}$

---

[2]We would like to mention that in these models, the determination of brake light status may be not consistent with reality. The so called "brake light status" actually means a certain kind of status of vehicles.

Table 8. Parameter values of CD model taken Knospe et al. (2000).

| Parameters | $L_{cell}$ | $L_{veh}$ | $v_{max}$ | $p_b$ | $p_0$ | $p_d$ | $a$ | $b$ | $g_{safety}$ | $h$ |
|---|---|---|---|---|---|---|---|---|---|---|
| Units | m | $L_{cell}$ | $L_{cell}/s$ | - | - | - | $L_{cell}/s^2$ | $L_{cell}/s^2$ | $L_{cell}$ | s |
| Value | 1.5 | 5 | 20 | 0.94 | 0.5 | 0.1 | 1 | 1 | 7 | 6 |

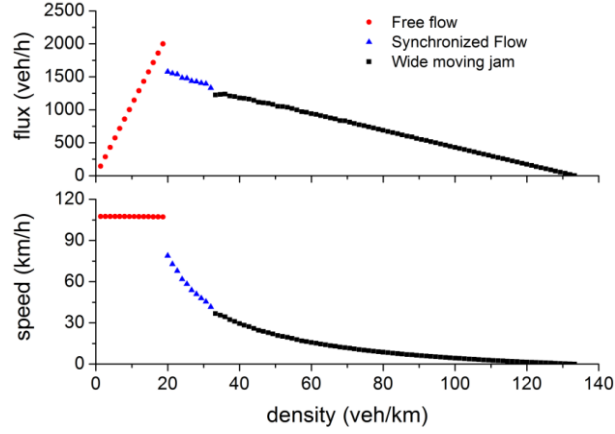

Fig. 20. The fundamental diagram of CD model on a circular road with length 7500m. The parameters are shown in Table 8.

Fig. 20 shows the fundamental diagram of the model. Three density ranges can be classified. In the synchronized flow branch, a phase separation phenomenon will occur and the system is the coexistence of free flow phase and congested flow phase (Fig. 21(*a*)). Note that the congested flow consists of both synchronized flow and jams. In other words, the model cannot correctly reproduce the synchronized flow. With the increase of the density, the free flow region shrinks (see Fig. 21(*a*) and (b)). In the wide moving jam branch, the free flow region disappears and only the congested flow exists (Fig. 21(c)). If one continues to increase the density, the jams will gradually invade the synchronized flow (Fig. 21(c) and (d)). Unfortunately, the metastable states, the F→S and S→J transitions cannot be simulated by CD model.

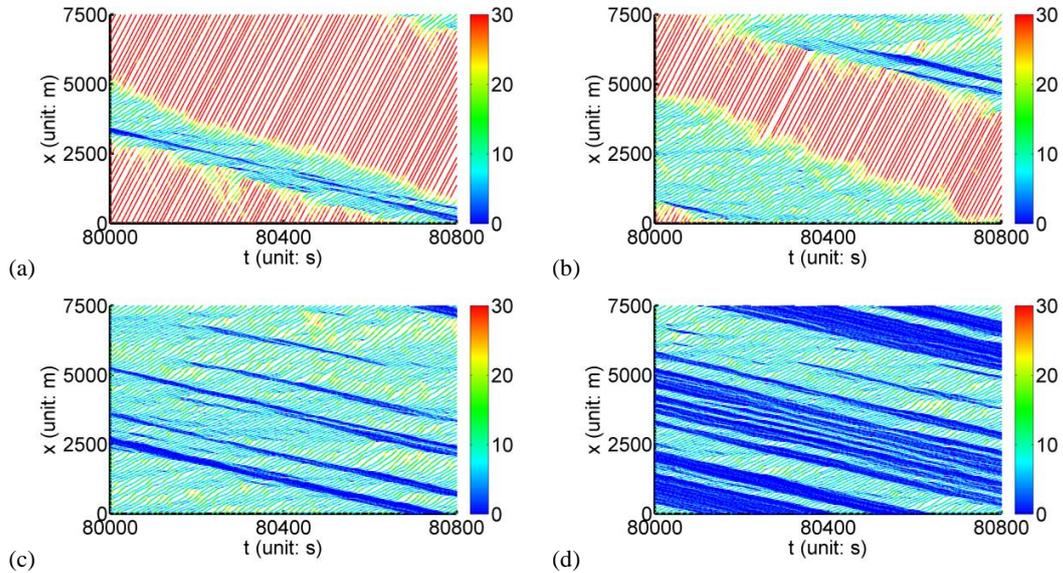

Fig. 21. The spatiotemporal diagram of the CD model on the circular road. (a) $\rho$=23veh/km; (b) $\rho$=28veh/km; (c)

### 3.3.2 The modified comfortable driving model

Jiang and Wu (2003) proposed a modified comfortable driving (MCD) model to reproduce both light and heavy synchronized flows. In this model, the formulations of $\hat{a}$, $\hat{d}$, $\hat{v}$, $\hat{v}_r$, and $\hat{p}$ are as follows.

(i) $\hat{a} = \begin{cases} a_1, & \text{if } (s_l = 0 \text{ or } t_h \geq t_s) \text{ and } v > 0, \\ a_2, & \text{if } v = 0, \\ 0, & \text{otherwise.} \end{cases}$

(ii) $\hat{d} = d_n + \max(v_{anti} - g_{safety}, 0)$.

(iii) $\hat{v} = v_{max}$.

(iv) $\hat{v}_r = b$.

(v) $\hat{p} = \begin{cases} p_b, & \text{if } s_l = 1 \text{ and } t_h < t_s, \\ p_0, & \text{if } v = 0 \text{ and } t_{st} \geq t_c, \\ p_d, & \text{otherwise.} \end{cases}$

In step (i), the acceleration of a stopped car is assumed to be $a_2$ and that of a moving car is $a_1$. Note that sometimes the vehicles do not accelerate in step (i). Moreover, $b \geq a_2$ should be satisfied to simulate the wide moving jams. The brake light state $s'$ in the next time step is determined as follows in the MCD model, which is different from that in the CD model,

$$s' = \begin{cases} 1, & \text{if } v' < v, \\ 0, & \text{if } v' > v, \\ s, & \text{otherwise.} \end{cases}$$

Fig. 22 shows the fundamental diagram of the MCD model. One can see that the results are similar to the Gao model. The F→S transition cannot be reproduced.

Table 9. Parameter values of MCD model taken from Jiang and Wu (2003).

| Parameters | $L_{cell}$ | $L_{veh}$ | $v_{max}$ | $p_b$ | $p_0$ | $p_d$ | $b$ | $a_1$ | $a_2$ | $g_{safety}$ | $h$ | $t_c$ |
|---|---|---|---|---|---|---|---|---|---|---|---|---|
| Units | m | $L_{cell}$ | $L_{cell}/s$ | - | - | - | $L_{cell}/s^2$ | $L_{cell}/s^2$ | $L_{cell}/s^2$ | $L_{cell}$ | s | s |
| Value | 1.5 | 5 | 20 | 0.94 | 0.5 | 0.1 | 1 | 2 | 1 | 7 | 6 | 7 |

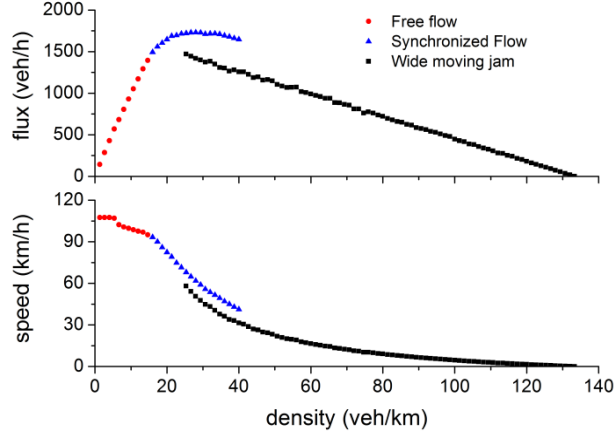

Fig. 22. The fundamental diagram of MCD model on the circular road. The parameters are shown in Table 9.

To overcome this deficiency, Jiang and Wu (2005) further introduced a new variable $t_{free}$. It denotes the time that car $n$ is in the state $v \geq v_c$. It is supposed that if $v' \geq v_c$ and $t_{free} \geq t_{c1}$, then $s$ remains to be zero despite of the value of $v$. For the determination of $t_{free}$, it is simply that

$$t'_{free} = \begin{cases} t_{free} + 1, & \text{if } v' \geq v_c, \\ 0, & \text{otherwise.} \end{cases}$$

Here $v_c$ and $t_{c1}$ are parameters and $t'_{free}$ is the $t_{free}$ at the next time step.

Simulation shows that the improved MCD model can simulate the F→S transition, see the fundamental diagram in Fig. 23 and spatiotemporal evolution pattern in Fig. 24(a). Note that in the improved MCD model, the synchronized flow does not coexist with the free flow, see Fig. 24(a) and Fig. 25(a). Nevertheless, the model has the same deficiency as the improved model of Gao et al., i.e., the velocity fluctuates too much in synchronized flow, see Fig. 25(b).

Table 10. Parameter values of improved MCD model taken from Jiang and Wu (2005).

| Parameters | $L_{cell}$ | $L_{veh}$ | $v_{max}$ | $p_b$ | $p_0$ | $p_d$ | $b$ |
|---|---|---|---|---|---|---|---|
| Units | $m$ | $L_{cell}$ | $L_{cell}/s$ | - | - | - | $L_{cell}/s^2$ |
| Value | 1.5 | 5 | 20 | 0.94 | 0.5 | 0.1 | 1 |
| Parameters | $a_1$ | $a_2$ | $g_{safety}$ | $h$ | $v_c$ | $t_c$ | $t_{c1}$ |
| Units | $L_{cell}/s^2$ | $L_{cell}/s^2$ | $L_{cell}$ | s | $L_{cell}/s$ | s | s |
| Value | 2 | 1 | 7 | 6 | 18 | 10 | 30 |

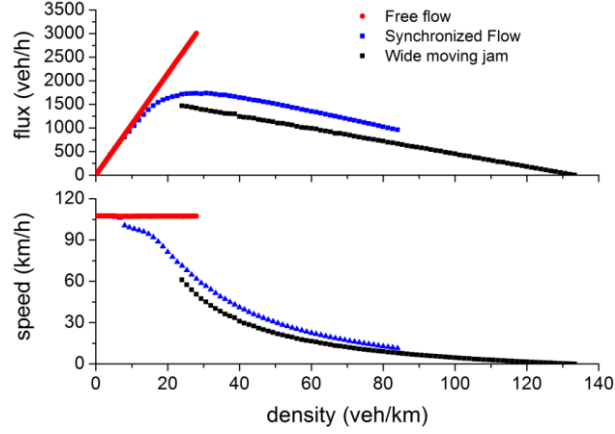

Fig. 23. The fundamental diagram of improved MCD model on the circular road with length 7500m. The parameters are shown in Table 10.

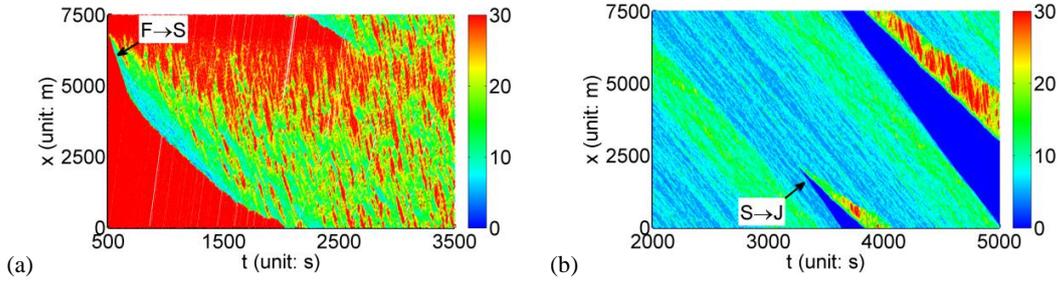

Fig. 24. The simulation results of the spatiotemporal diagram. (a) $\rho$=29veh/km; (b) $\rho$=85veh/km.

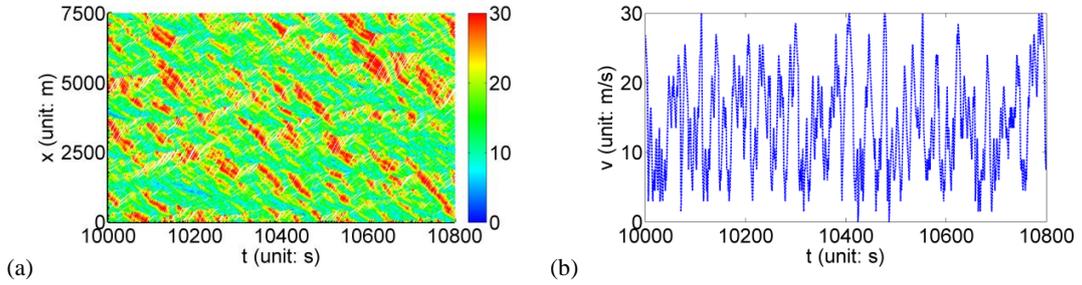

Fig. 25. The simulation results of spatiotemporal diagram. (a) $\rho$=33veh/km; (b) The time series of velocity of a single vehicle.

### *3.3.3 The desire time gap brake light model*

Tian et al. (2017) found that in the comfortable driving model, when the leading vehicle decelerates sharply, there is not enough time for following vehicle to smoothly adapt its speed due to the reaction delay. As a result, jam emerges, and synchronized flow cannot be maintained.

To overcome this deficiency, they have revised the braking rule by introducing a desired time gap larger than 1 second, named as the desire time gap brake light (DTGBL) model. In DTGBL model, the formulations of $\hat{a}$, $\hat{d}$, $\hat{v}$, $\hat{v}_r$, and $\hat{p}$ are as follows.

(i) $\hat{a} = \begin{cases} a_1, & \text{if } (s_l = 0 \text{ or } t_h \geq t_s) \text{ and } v > 0 \\ a_2, & \text{otherwise} \end{cases}$

(ii) $\hat{d} = \lceil (d_n + \max(v_{anti} - g_{safety}, 0))/T \rceil$

(iii) $\hat{v} = v_{max}$

(iv) $\hat{v}_r = b$

(v) $\hat{p} = \begin{cases} p_b, & \text{if } s_l = 1 \text{ and } t_h < t_s, \\ p_0, & \text{if } v = 0, \\ p_d, & \text{otherwise.} \end{cases}$

where the brake light rule in the CD model is applied. Here $\lceil x \rceil$ returns the minimum integer no smaller than $x$.

Compared with the CD model, the highlight feature is that the desired time gap is set as T > 1. Fig. 26 shows the fundamental diagram of this model, which can reproduce the free flow, the synchronized flow, the jam as well as F→S (Fig. 27(a)), S→J transition (Fig. 27(b)). Also note that the fluctuation of speed in synchronized flow is reasonable, see Fig. 28.

Table 11. Parameter values of DTGBL model taken from Tian et al. (2017).

| Parameters | $L_{cell}$ | $L_{veh}$ | $v_{max}$ | T | $p_b$ | $p_0$ | $p_d$ | $g_{safety}$ | h | $a_1$ | $a_2$ | b |
|---|---|---|---|---|---|---|---|---|---|---|---|---|
| Units | m | $L_{cell}$ | $L_{cell}$/s | s | - | - | - | $L_{cell}$ | s | $L_{cell}/s^2$ | $L_{cell}/s^2$ | $L_{cell}/s^2$ |
| Value | 1.5 | 5 | 20 | 1.8 | 0.94 | 0.5 | 0.1 | 7 | 6 | 2 | 1 | 1 |

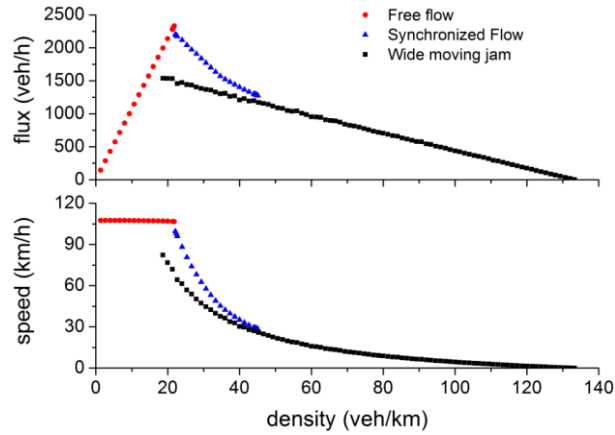

Fig. 26. The fundamental diagram of DTGBL model with length 3000m. The parameters are shown in Table 11.

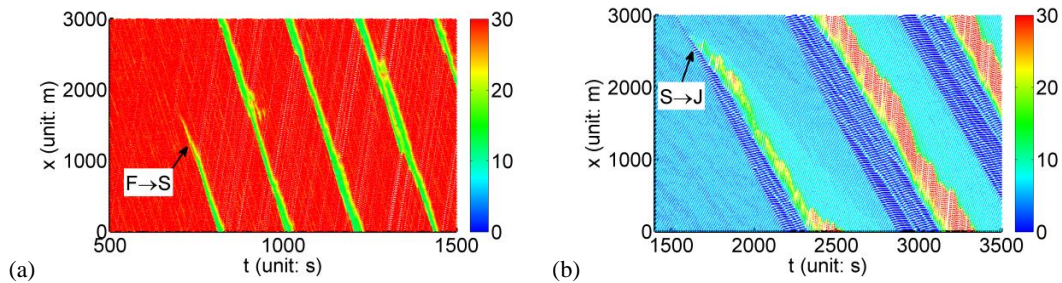

Fig. 27. The spatiotemporal pattern of (a) $\rho$ =22veh/km, describes F→S, (b) $\rho$=67veh/km, describes S→J transition.

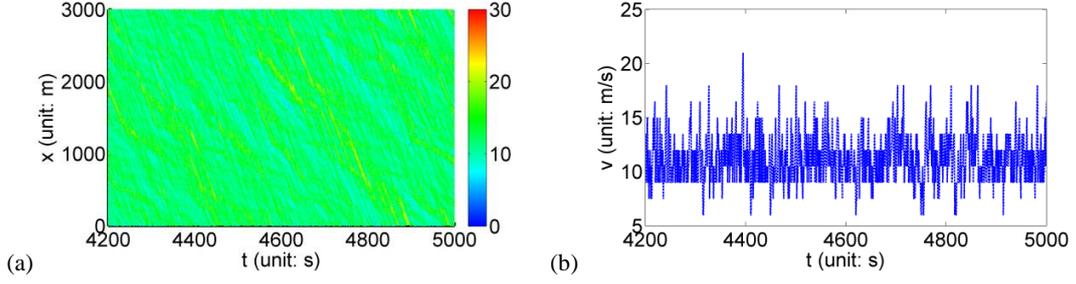

Fig. 28. The simulation results of spatiotemporal diagram. (a) $\rho$=37veh/km; (b) The time series of velocity of a single vehicle.

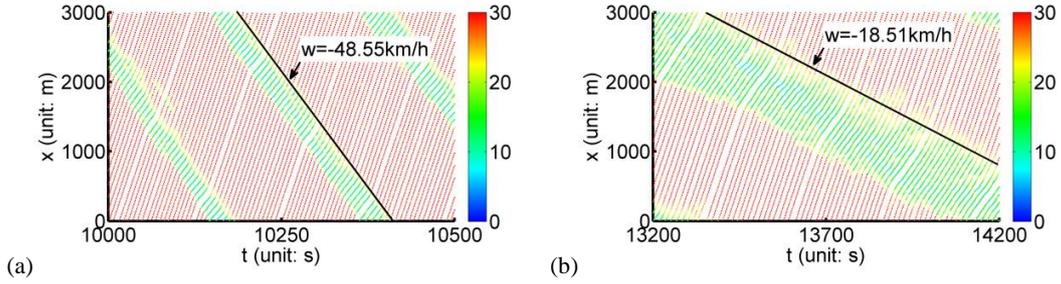

Fig. 29. The spatiotemporal pattern of (a) DTGBL model with $\rho$=24veh/km, (b) the improved DTGBL model with $\rho$=24veh/km. The parameters of improved DTGBL model are shown in Table 12.

However, the propagation speed of synchronized flow is too fast in this model, see Fig. 29(a). To overcome this shortcoming, Wang et al. (2017) assumes that: (i) the drivers would be sensitive to the brake light only when their speeds are larger than a critical speed $v_{cri}$; (ii) The effectiveness of the anticipation $g_{safety}$ increases with the increase of the speed of the following vehicle, i.e. $g_{safety} = \lceil b + g_s v / v_{max} \rceil$. They have modified DTGBL model by the reformulations of $\hat{a}$ and $\hat{p}$.

(i) $\hat{a}=a$

(v) $\hat{p} = \begin{cases} p_b, & \text{if } s_l = 1 \text{ and } t_h < t_{sa} \text{ and } v > v_{cri}, \\ p_0, & \text{if } v = 0, \\ p_d, & \text{otherwise.} \end{cases}$

The propagation speed of synchronized flow is consistent with the reality, see Fig. 29(b). Note that in the DTGBL model, synchronized flow can coexist with the free flow.

Table 12. Parameter values of improved DTGBL model taken from Wang et al. (2017).

| Parameters | $L_{cell}$ | $L_{veh}$ | $v_{max}$ | $T$ | $p_b$ | $p_0$ | $p_d$ | $h$ | $a$ | $b$ | $g_s$ | $v_{cri}$ |
|---|---|---|---|---|---|---|---|---|---|---|---|---|
| Units | m | $L_{cell}$ | $L_{cell}$/s | s | - | - | - | s | $L_{cell}$/s$^2$ | $L_{cell}$/s$^2$ | $L_{cell}$/s$^2$ | $L_{cell}$/s |
| Value | 1.5 | 5 | 20 | 2.0 | 0.94 | 0.45 | 0.1 | 6 | 1 | 1 | 5 | 9 |

*3.4 The limited deceleration model*

In the previous models, vehicles have infinite deceleration capability and can decelerate from any speed to zero speed in a time step. In some models, the limited deceleration capability of vehicles has been considered (Lee et al., 2004; Pottmeier et al., 2007; Lárraga and Alvarez-Icaza, 2010; Jin and Wang, 2011; Kokubo et al. 2011; Chmura et al., 2014).

One example of this type of models was proposed by Lee et al. (2004). Optimistic and pessimistic driving are introduced. Drivers evaluate the local situation to decide whether they will drive optimistically ($\gamma = 0$) or pessimistically ($\gamma = 1$). The parallel updating rules are as follow:

Step 1: Determine the safe velocity $v_s$:

$$v_s = \max\left(c \mid x + \Delta + \sum_{i=0}^{\tau(c)}(c - bi) \leq x' + \sum_{i=1}^{\tau_l(v_l)}(v_l - bi)\right).$$

Step 2: Determine the randomization probability $p$:

$$p = \max\left(p_d, p_0 - (p_0 - p_d)v/v_{slow}\right).$$

Step 3: Velocity update:

$$\tilde{v} = \min\left(v_{max}, v + a, \max(0, v - b, v_s)\right).$$

Step 4: Randomization:

$$v' = \begin{cases} \max(0, v - b, \tilde{v} - a), & \text{if } r < p, \\ \max(0, v - b, \tilde{v}) & \text{otherwise.} \end{cases}$$

Step 5: Vehicle Movement:
$$x' = x + v'.$$

where,

$$\gamma = \begin{cases} 0, & \text{if } v \leq v_l \leq v_{l+1} \text{ or } v_{l+1} \geq v_{fast}, \\ 1, & \text{otherwise.} \end{cases}$$

$$\Delta = L_{veh} + \gamma \max\left(0, \min(g_{add}, v - g_{add})\right).$$

$$\tau(v) = \gamma v/b + (1-\gamma)\max\left(0, \min(v/b, t_{safe}) - 1\right).$$

$$\tau_l(v) = \gamma v/b + (1-\gamma)\min(v/b, t_{safe}).$$

Here, $a$ and $b$ are restricted acceleration and deceleration. $\gamma$ denotes the driving condition where 0 represents a relative safe condition and 1 represents a relative tense condition. $\tau(v)$ and $\tau_l(v)$ respectively are the time from deceleration to stop of the followed vehicle and the leading one. $c$ is the safe velocity value of the follower, and its maximum $v_s$ denotes moving as fast as possible while keeping safe. $v_{fast}$ is a high speed slightly below $v_{max}$. $v \leq v_l \leq v_{l+1}$ means the drivers drive in a platoon that are driving equally fast or faster than themselves, where $v_{l+1}$ is the speed of next vehicles of their leaders. $\Delta$ is the minimal coordinate difference required by the follower to guarantee its safety. $g_{add}$ is introduced for an additional security gap in the defensive state and $t_{safe}$ is a maximal time step during which the follower observes his/her own safety in the optimistic state.

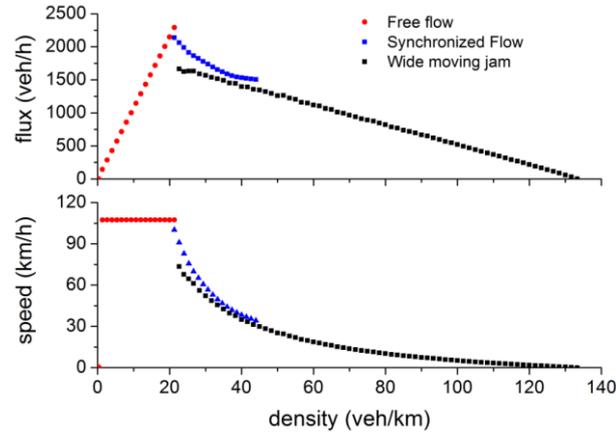

Fig. 30. The fundamental diagram of the Lee model on the circular road with length 3000m. The parameters are shown in Table 13.

Table 13. Parameter values of Lee model taken from Lee et al. (2004).

| Parameters | $L_{cell}$ | $L_{veh}$ | $v_{max}$ | $a$ | $b$ | $v_{fast}$ | $t_{safe}$ | $g_{add}$ | $p_0$ | $p_d$ | $v_{slow}$ |
|---|---|---|---|---|---|---|---|---|---|---|---|
| Units | m | $L_{cell}$ | $L_{cell}/s$ | $L_{cell}/s^2$ | $L_{cell}/s^2$ | $L_{cell}/s$ | s | $L_{cell}$ | - | - | $L_{cell}/s$ |
| Value | 1.5 | 5 | 20 | 1 | 2 | 19 | 3 | 4 | 0.32 | 0.11 | 5 |

Fig. 30 shows the fundamental diagram, which can reproduce the free flow, the synchronized flow, the jam as well as F→S (Fig. 31(a)), S→J transition (Fig. 31(b)). Also note that the synchronized flow does not coexist with free flow and the synchronized flow is quite homogeneous, see Fig. 32.

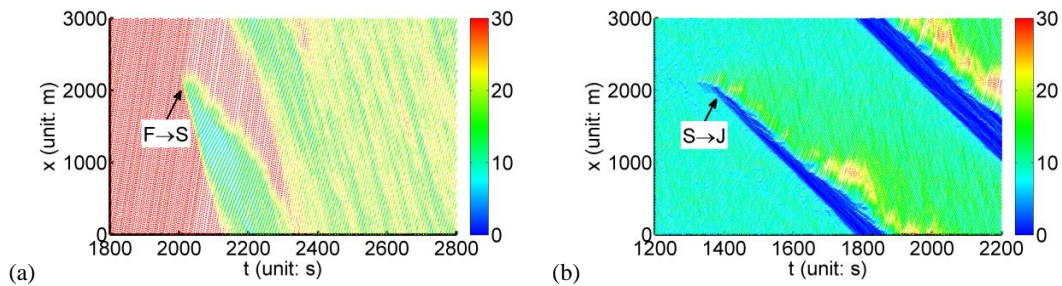

Fig. 31. The spatiotemporal diagrams of Lee model. (a) $\rho$=27veh/km; (b) $\rho$=50veh/km.

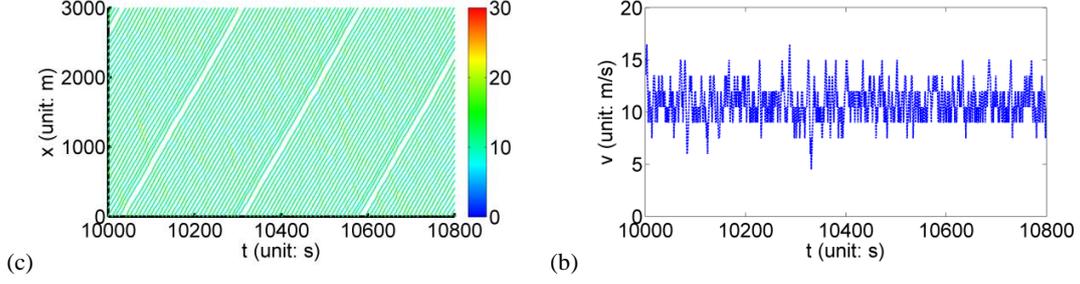

Fig. 32. The simulation results of spatiotemporal diagram. (a) $\rho$=40veh/km; (b) The time series of velocity of a single vehicle.

Pottmeier et al. (2007) pointed out that accidents would happen due to the delayed change from the optimistic to the pessimistic state in the model of Lee et al. Vehicle may react too late to changes around the leading vehicles, especially of the second-nearest car ahead. They showed that accident may happen (i) if all three vehicles involved in the calculation of $\gamma$ drive optimistically with the same velocity; (ii) if the velocity difference between $v$ and $v_l$ is too high. Thus, they revised the definition of $\gamma$ as follows,

$$\gamma = \begin{cases} 0, & \text{if } v \leq v_l < v_{l+1} \text{ or} \\ & (v_{l+1} \geq v_{\text{fast}} \text{ and } v - v_l \leq b \text{ and } s_{l+1} = 0), \\ 1, & \text{otherwise.} \end{cases}$$

To prevent the first type accident, the inequality $v \leq v_l \leq v_{l+1}$ is changed into $v \leq v_l < v_{l+1}$. An upper limit of the difference between $v$ and $v_l$ is adopted to avoid the accidents due to a large velocity difference. Moreover, a stronger connection between each vehicle and its next nearest leader $l+1$ is added by the introduction of the indicator $s$, which denotes whether the vehicle has reduced its velocity because of its surrounding, but not because of dawdling (i.e. due to randomization) so that the follower is able to sense a critical situation early enough.

$$s = \begin{cases} 1, & \text{if } \tilde{v} < v, \\ 0, & \text{otherwise.} \end{cases}$$

The improved model overcomes the "accident" problem in the model of Lee et al., and its other macroscopic and microscopic results are almost not influenced.

## 4. Conclusion and outlook

*4.1 Conclusions*

In recent two decades, "synchronized flow" is perhaps one of the most important concepts and findings in traffic flow studies. Many models were proposed to reproduce the synchronized flow as well as traffic breakdown from free flow to synchronized flow. In the CA models, the randomness is an important component. Thus, the stochasticity is naturally considered, which is found to play a nontrivial role in traffic flow (Laval et al., 2014, Jiang et al., 2014, 2015, 2018, Treiber and Kesting, 2017). If the stochasticity is removed from CA models, none of them can show the simulation results discussed in previous sections. Moreover, the time step is usually set to one second while it is usually 0.1 second or even smaller in car-following models. As a result, the simulation time is much less in CA models and suitable for large scale traffic network.

In this paper, we have reviewed the CA models that reproduce synchronized. One main contribution is that we classified the models into a few different types, concerning their main characteristics. The models proposed by Kerner and his colleagues took the speed adaptation into account explicitly. The second type is generalized NaSch model. The class of brake light models also belongs to the generalized NaSch models. The difference from the second type is that the status of brake light is considered. the In the fourth type of models, the limit deceleration capacity of vehicles is introduced.

*4.2 Future Directions*

While all these models can reproduce the synchronized flow, some common deficiencies exist for almost all models. (1) In real traffic flow, with the increase of the density, the free flow branch bends downward. The lowest speed in the free flow can be as low as 70-80 km/h. However, in the CA models, the free flow branch is almost a straight line. The lowest speed in free flow is very close to the maximum speed. Therefore, how to capture the feature of free flow branch is a common task for CA modeling. (2) Three-phase traffic flow models usually have many parameters and most of them have no direct physical meanings and thus cannot be measured. The simple model needs to be developed. (3) The rules of many models are usually based on the behavioral assumptions. The data validation and justification are absent. (4) A quantitative comparison with real traffic data is lacked for most CA models. (5) The extensions of the single lane models to multi-lane models and further to the traffic networks are needed.

Moreover, due to a lack of precise traffic data, there are many other unclear issues. For example, (i) whether the synchronized flow could coexist with free flow; (ii) how does the traffic flow oscillate and what is the oscillation strength in the synchronized flow; (iii) how does the free flow spontaneously transit into synchronized flow on a road without bottleneck. To reveal the issues, significant efforts are still needed in the future work toward traffic flow experiment and observation, data collection and analysis.

**Acknowledgements**

JFT was supported by the National Natural Science Foundation of China (Grant No. 71771168). RJ was supported by the Natural Science Foundation of China (Grant No. 71621001).